\documentclass[aps,prd,superscriptaddress,notitlepage,nofootinbib,10pt]{revtex4-2}

\usepackage{amsmath}
\usepackage{amssymb}
\usepackage{enumitem}
\usepackage{mmacells}
\usepackage{hyperref}
\usepackage{alphabeta}

\DeclareRobustCommand{\textneq}{%
  \begingroup
  \ooalign{\hidewidth/\hidewidth\cr=\cr}\vphantom{/}%
  \endgroup
}

\newcommand\mmaArrow{\raisebox{-.2ex}{-\kern-.5ex\textrightarrow}}
\newcommand\mmaColonArrow{\raisebox{.2ex}{:}\kern-.4ex\textrightarrow\kern-.40ex}
\newcommand\mmaEq{\raisebox{.2ex}{:}\kern-.2ex=\kern-.40ex}
\newcommand\mmaCurly[1]{\(\{\)#1\(\}\)}
\newcommand\mmaOtimes{\(\boldsymbol{\otimes}\)}
\newcommand\mmaTimes{\(\boldsymbol{\times}\)}
\newcommand\mmaOplus{\(\boldsymbol{\oplus}\)}
\newcommand\mmaWedge{\(\boldsymbol{\wedge}\)}
\newcommand\mmaEquiv{\(\boldsymbol{\equiv}\)}

\newcommand\mmaEth{\(\eth\)}
\newcommand\mmaHbar{\(\hbar\)}
\newcommand\mmaMu{\textit{\mu}}
\newcommand\mmaNu{\textit{\nu}}
\newcommand\mmaAlpha{\textit{\alpha}}
\newcommand\mmaBeta{\textit{\beta}}
\newcommand\mmaGamma{\textit{\gamma}}
\newcommand\mmaDelta{\textit{\delta}}
\newcommand\mmaLambda{\textit{\lambda}}
\newcommand\mmaRho{\textit{\rho}}
\newcommand\mmaSigma{\textit{\sigma}}
\newcommand\mmaEpsilon{\textit{\epsilon}}
\newcommand\mmaPsi{\textit{\psi}}
\newcommand\mmaTau{\textit{\tau}}
\newcommand\mmaPhi{\(\boldsymbol{\phi}\)}
\newcommand\mmaVarphi{\(\boldsymbol{\varphi}\)}

\begin{document}

\title{SimpleTensor - a user-friendly Mathematica package for elementary tensor and differential-geometric calculations}

\author{D.~O.~Rybalka}
\affiliation{Max Planck Institute for the Physics of Complex Systems, Nöthnitzer Str. 38, 01187 Dresden, Germany}

\date{\today}

\begin{abstract}
  In this paper we present a short overview of the new Wolfram Mathematica package intended for elementary "in-basis" tensor and differential-geometric calculations.
  In contrast to alternatives our package is designed to be easy-to-use, short, all-purpose, and hackable.
  It supports tensor contractions using Einstein notation, transformations between different bases, tensor derivative operator, expansion in basis vectors and forms, exterior derivative, and interior product.
\end{abstract}

\maketitle

\section{Introduction}
As a motivation for this project can serve a simple expression from electrodynamics: the dual electromagnetic tensor $\frac{1}{2}\epsilon^{\mu\nu\alpha\beta}\partial_{\alpha}A_{\beta}$.
To the best of our knowledge there is no easy way to calculate this expression in base Mathematica without shuffling explicitly index contractions and paying close attention to metric signatures.
Furthermore, the search for an external package proved to be unsatisfactory as well, with the results being hard to use and containing large complexity overhead for such a simple task.
Therefore, we created a user-friendly tensor computation package for situations, where clarity is more important than performance.

After 12 versions and 33 years of development Wolfram Mathematica is an invaluable tool for both scientists and students.
Its functionality covers most needs in mathematical calculations starting from undergraduate courses up to physics-theoretical research.
The most glaring exceptions in its scope is the tensor calculus (which is present only in the form of explicit lists manipulation) and differential geometry.

The reason for that may be the fact that these topics cannot be fitted well in Mathematica's paradigm of "stateless" computations, i.e., the idea that each function must use only the arguments explicitly provided to it. 
And while this approach works well in most cases, in tensor calculus and differential geometry it becomes burdensome. 
For example, a simple contraction of tensors require not only the actual tensor components, but also dimensions and often metrics of spaces where the corresponding indices are defined.

There already exist a few packages intended to provide this functionality, for example, xAct \cite{xAct}, TensoriaCalc \cite{TensoriaCalc}, grt \cite{grt}, Ricci \cite{Ricci} just to name a few.
However, in our opinion, they all suffer from one or more of the following flaws:
\begin{itemize}[topsep=6pt,itemsep=0pt,parsep=4pt]
  \item The package is old, outdated, or not supported anymore;
  \item The package is hard to master (or even to grasp), as it contains dozens or even hundreds of public functions;
  \item The package is focused primarily on the General Relativity computations;
  \item The package is hard to adapt for your personal needs.
\end{itemize}

With this paper we introduce SimpleTensor - a yet-another package for elementary tensor and differential-geometric calculations, intended to fill the gap in base Mathematica functionality.
As the name suggests, it differs from the alternatives by being accessible, concise (only 11 public functions), up-to-date, all-purpose, and well-documented (hackable).
It also provides basis for elementary in-basis differential-geometric computations. 

However, some sacrifices had to be made in order to achieve these goals. 
SimpleTensor is not optimized for any large-scale computations and will be slower than the alternatives.
It was also not specialized for any particular area of physics or mathematics and so loses considerably on functionality for specialized computations.
Nevertheless, even despite these flaws we hope that SimpleTensor can find its auditory among both students and scientists.

The latest version of SimpleTensor can be found in the author's \href{https://github.com/drybalka/SimpleTensor}{GitHub repository} \cite{Repository}.
Located there are the source code, the generated \texttt{.m} package file, together with the tests and the example session files.
The tests file can be used to get a quick overview of all the capabilities of SimpleTensor as well as to test custom modifications of the source code (enhances hackability of the package).
The example session will be examined in detail later in this paper.

In the next chapter we present the technical information concerning SimpleTensor, discuss the package design and document provided public functions.

\section{Technical information}
The package installation process is similar to the most other external Mathematica packages.
One may simply copy the \texttt{SimpleTensor.m} file to the Applications folder in the \texttt{\$UserBaseDirectory}.
The package then can be loaded as any other Mathematica package with \texttt{Needs} command:
\begin{mmaCell}{Input}
  Needs["SimpleTensor`"]
\end{mmaCell}
Alternatively, if any source code modifications are desirable, one may simply edit and execute the contents of \texttt{SimpleTensor.nb} file.

Let us now give a broad overview of the package.
The main actors of SimpleTensor are tensors, which have to be explicitly defined by their components in some coordinate system in some space.
Spaces are specified by a unique name, dimension, indices, and coordinates.
Both indices and coordinates are the usual Mathematica symbols, that have to be unique among all spaces to avoid name collisions.
Indices can be either contravariant (upper) - denoted with a positive sign, e.g., \texttt{+\mmaMu} or simply \texttt{\mmaMu}, or covariant (lower) - denoted with a negative sign, e.g., \texttt{-\mmaMu}.
In tensors the indices are then denoted in square brackets, so that $F^\mu_{\;\;\nu}$ becomes \texttt{F[\mmaMu,-\mmaNu]}.

Tensor components can be assigned using the \texttt{SetTensor} command \texttt{\mmaEquiv} operator).
The main command of SimpleTensor is \texttt{GetArray}, which performs all tensor calculations on an expression and returns the resulting array.
Special meaning is reserved for a few tensor names: \texttt{metric} is used to lower and raise indices as well as for basis change, \texttt{basis} is a tensor of differential-geometric basis forms/vectors, and \texttt{\mmaEth} is the tensor derivative.
Basic differential geometric operations are provided by \texttt{DirectProduct} (\mmaOtimes), \texttt{WedgeProduct} (\mmaWedge), \texttt{ExteriorDerivative}, and \texttt{InteriorProduct}.
Below we document the public methods of SimpleTensor in more detail.

\begin{description}[topsep=6pt,itemsep=4pt,parsep=4pt]
  \item[\texttt{DefineSpace[name,\,dim]}]  defines a space of dimension \texttt{dim} with a string \texttt{name} on which tensors live.
    The optional variables are:
    \begin{itemize}[topsep=0pt,itemsep=0pt,parsep=2pt,label=-]
      \item \texttt{index} - a list of indices on this space or a string prefix to generate names automatically; if missing, name is used as a prefix.
      \item \texttt{coord} - a list of coordinates on this space or a string prefix to generate names automatically; if missing, name is used as a prefix.
    \end{itemize}
    For example, we can define a 4-dimensional Minkowski space with \texttt{DefineSpace["Minkowski",\;4,\;index \mmaArrow\,\mmaCurly{\mmaMu,\mmaNu,\mmaAlpha,\mmaBeta},\;coord\,\mmaArrow\,\mmaCurly{t,x,y,z}]}.
    This will additionally define a set of basis vectors \texttt{\mmaEth\,\!t,\mmaEth\,\!x,\mmaEth\,\!y,\mmaEth\,\!z} and a set of basis forms \texttt{dt,dx,dy,dz}.

    If \texttt{index} or \texttt{coord} optional arguments are given as a string then it is used as a prefix.
    For example, \texttt{index\,\mmaArrow\,"L"} is equivalent to \texttt{index\,\mmaArrow\,\mmaCurly{iL0,iL1,...}} and \texttt{coord\,\mmaArrow\,"L"} to \texttt{coord\,\mmaArrow\,\mmaCurly{xL0,xL1,...}}.
    If \texttt{index} or \texttt{coord} are not specified then the space \texttt{name} is used as a prefix.

    Defining a space for a second time deletes all definitions connected to the old one.
    Thus, one can effectively delete a space by redefining it with zero dimension.

  \item[\texttt{SetTensor[t[\mmaMu],\,arr]}] associates array \texttt{arr} to a tensor \texttt{t[\mmaMu]}.
    For convenience this function overrides the built-in \texttt{Congruent} (\mmaEquiv) function.
    To change the default behavior edit corresponding lines in the package file.

    Naturally, any tensor name (head) and upper/lower indices can be used here.
    Indices from the same space can be used interchangeably.
    For example, in the Minkowski space defined above \texttt{t[\mmaMu,-\mmaNu]}, \texttt{t[\mmaNu,-\mmaMu]}, \texttt{t[\mmaAlpha,-\mmaBeta]}, and even \texttt{t[\mmaMu,-\mmaMu]} inside \texttt{SetTensor} mean the same.

    The dimension compatibility of \texttt{arr} and indices must be correct and will be checked on assignment.
    It is possible, however, to have extra dimensions in \texttt{arr}, in which case tensor elements are assumed to be matrices.
    We will demonstrate in the next chapter how this allows to define a tensor of Dirac matrices $\gamma^\mu$.

    Note, that assignments for the same tensor head but different signatures or spaces are independent and do not overwrite each other.
    This allows for simpler expressions (especially if \texttt{metric} is complicated), but the burden of ensuring consistency in this case falls on user's shoulders.

  \item[\texttt{GetArray[expr]}] substitutes tensor variables with arrays and performs contractions and math operations.
    The optional variables are:
    \begin{itemize}[topsep=0pt,itemsep=0pt,parsep=2pt,label=-]
      \item \texttt{indices} - a list of indices in which order the tensor should be transposed. If omitted \texttt{indices} are taken to be the free indices in \texttt{expr} in standard order, i.e., the one returned by the standard \texttt{Sort[]} function.
    \end{itemize}
    Be mindful of the standard ordering as \texttt{GetArray[F[\mmaMu,\mmaNu]]\,\textneq\,GetArray[F[\mmaNu,\mmaMu]]} in general.
    For example, for antisymmetric tensors \texttt{GetArray[F[\mmaMu,\mmaNu]\,+\,F[\mmaNu,\mmaMu]]\,==\,0} as expected.
    However, they are equal if one specifies the desired indices explicitly \texttt{GetArray[F[\mmaMu,\mmaNu]]\,==\,GetArray[F[\mmaMu,\mmaNu],\mmaCurly{\mmaMu,\mmaNu}]\,==\,GetArray[F[\mmaNu,\mmaMu],\mmaCurly{\mmaNu,\mmaMu}]}

  \item[\texttt{metric[-i1,\,-i2]}] is the metric tensor for the space in which \texttt{i1} and \texttt{i2} are defined.
    \texttt{metric[i1,i2]} is the inverse metric tensor.
    If \texttt{i1} and \texttt{i2} belong to different spaces then it works as a tetrad (vielbein) tensor and represents the change of basis.

    The \texttt{metric} tensor is automatically used by \texttt{GetArray} function to transform indices between different signatures and spaces.
    Multiple metrics will be chained where needed for complex space-signature transformations.

    The inverse \texttt{metric} will be automatically calculated on the assignment unless it was already explicitly assigned earlier.
    This allows for simpler expressions for complicated metrics, but consistency needs to be insured by the user.

  \item[\texttt{basis[i]}] is the tensor of basis forms for the space in which index \texttt{i} is defined.
    \texttt{basis[-i]} is the tensor of basis vectors.

    The \texttt{basis} tensor is protected and may not be modified by the user.
    Its value is automatically generated from coordinates during the \texttt{DefineSpace} call.
    Note, that \texttt{basis[i]} and \texttt{basis[-i]} are essentially different and are not connected through \texttt{metric}.

  \item[\texttt{\mmaEth[\textpm i]}] is the tensor derivative for the space in which index \texttt{i} is defined.
    \texttt{\mmaEth\textit{space}[\textpm i]} is also defined for a derivative in the space \texttt{\textit{space}}, which can be used with any \texttt{i}.

    In the Minkowski space example the components of \texttt{\mmaEth[-\mmaMu]} are $(\partial_t,\partial_x,\partial_y,\partial_z)$.
    The argument of the derivative must be given in square brackets after the tensor \texttt{\mmaEth[-\mmaMu][arg]} and may include other tensors.
    Note, that \texttt{\mmaEth[-\mmaMu]\,\textneq\,metric[-\mmaMu,a]\,\mmaEth[-a]} if \texttt{\mmaMu} and \texttt{a} are from different spaces.
    However, the space-specific derivatives satisfy this equality for any indices.

    Tensor names starting with \texttt{\mmaEth} are reserved for tensor derivatives and may not be modified.
    Their values are automatically generated from coordinates during the \texttt{DefineSpace} call.

  \item[\texttt{TensorPlus[args]}] is used to add matrix-valued tensors (for example, the set of Dirac matrices $\gamma^\mu$) with explicit lists (for example, \texttt{IdentityMatrix[4]}).
    It is equivalent to \texttt{Plus[]} in most cases, except it prevents the automatic threading over lists.
    For convenience this function overrides the built-in \texttt{CirclePlus} (\mmaOplus) function.

  \item[\texttt{DirectProduct[args]}] represents the tensor product of basis vectors and forms.
    It expands \texttt{args} automatically until a sum of basis tensor products is achieved.
    For convenience this function overrides the built-in \texttt{CircleTimes} (\mmaOtimes) function.

  \item[\texttt{WedgeProduct[args]}] represents the wedge product of basis forms.
    It expands \texttt{args} automatically until a sum of basis wedge products is achieved.
    For convenience this function overrides the built-in \texttt{Wedge} (\mmaWedge) function.

    The \texttt{WedgeProduct} of basis forms is automatically sorted into canonical order respecting anticommutation rules.
    If a basis form is repeated inside \texttt{WedgeProduct} it is set to zero.
    Both \texttt{WedgeProduct} and \texttt{DirectProduct} also work with the \texttt{basis} tensor.

  \item[\texttt{ExteriorDerivative[space][expr]}] gives the exterior derivative of \texttt{expr} in \texttt{space}.
    If latter is a \texttt{Sequence} of spaces, the exterior derivative is taken in their product space.
    If \texttt{space} is empty, the exterior derivative is taken in the product space of all defined spaces.

  \item[\texttt{InteriorProduct[expr1,\,expr2]}] is the symmetric interior product of basis vectors and forms in \texttt{expr1} and \texttt{expr2}.
    The \texttt{DirectProduct}`s of basises are multiplied successively starting from the first position.

  Basis vectors and forms correspond to each other with respect to the \texttt{InteriorProduct}, so that for the Minkowski space example \texttt{InteriorProduct[dx,\,\mmaEth\,\!x]\,==\,1} and \texttt{GetArray\,@\,InteriorProduct[basis[-\mmaMu],\,basis[\mmaNu]]\linebreak\,==\,IdentityMatrix[4]}.
  The interior product of composed expressions is performed successively starting from the first position, so that \texttt{InteriorProduct[\mmaEth\,\!y\mmaOtimes\mmaEth\,\!x,\,dx\mmaWedge dy]\,==\,-\,InteriorProduct[\mmaEth\,\!x,\,dx]\,=\,-1}.
\end{description}

As one can see from the documentation above there are only 11 public functions in the whole package.
Moreover, if one is not interested in differential geometry then the last 4 functions are irrelevant and can be simply ignored.
In the next chapter we present a few illustrative examples of how this limited functionality can already be used for various physics computations.
The notebook session with these examples, as well as the test-file with edge cases can be found in the project repository \cite{Repository}.

\section{Examples}

\subsection{Electromagnetic tensor}
In the first part we examine the applications of SimpleTensor on the example of undergraduate-level relativistic electromagnetism.
The following material can be found in most textbooks on classical electrodynamics \cite{Jackson:1998nia,field-theory}.
In this section we will first derive the relations between electromagnetic potentials and fields.
We will then calculate the field part of Maxwell equations in 2 different ways: explicitly from the EM tensor and using exterior calculus from the potential 1-form.
Finally, we will obtain the field transformations under a boost using a moving reference frame.

We start by defining a Minkowski space with mostly negative metric.
\begin{mmaCell}[moredefined={DefineSpace, metric}]{Input}
  DefineSpace["Minkowski",\;4,\;index\,\mmaArrow\,\mmaCurly{\mmaUnd{\mmaMu},\mmaUnd{\mmaNu},\mmaUnd{\mmaLambda},\mmaUnd{\mmaRho}},\;coord\,\mmaArrow\,\mmaCurly{t,x,y,z}]
  metric[-\mmaUnd{\mmaMu},-\mmaUnd{\mmaNu}]\,\mmaEquiv\,DiagonalMatrix[\mmaCurly{1,-1,-1,-1}]
\end{mmaCell}
Next we set a Levi-Civita tensor $\epsilon^{\mu\nu\lambda\rho}$ and a time direction vector $T^{\mu}$.
The latter is done using the full function name as illustration.
\begin{mmaCell}[addtoindex=1,moredefined={SetTensor}]{Input}
  \mmaUnd{\mmaEpsilon}[\mmaUnd{\mmaMu},\mmaUnd{\mmaNu},\mmaUnd{\mmaLambda},\mmaUnd{\mmaRho}]\,\mmaEquiv\,Normal\,@\,LeviCivitaTensor[4]
  SetTensor[T[\mmaUnd{\mmaMu}],\,\mmaCurly{1,0,0,0}]
\end{mmaCell}
Finally, we set a general electromagnetic 4-potential $A^\mu$.
Note, that we use the same coordinates defined in the Minkowski space.
\begin{mmaCell}[addtoindex=1]{Input}
  A[\mmaUnd{\mmaMu}]\,\mmaEquiv\,\mmaCurly{\mmaUnd{\mmaPhi}[t,x,y,z],\,A1[t,x,y,z],\,A2[t,x,y,z],\,A3[t,x,y,z]}
\end{mmaCell}
Individual components of a tensor can be extracted by contracting it with a corresponding unit vector.
For example, by contracting EM 4-potential $A^\mu$ with a time direction vector $T^\mu$ we get the corresponding zeroth component - the scalar potential $\phi$.
Alternatively one can first calculate the array and then access its components.
\begin{mmaCell}[moredefined={GetArray}]{Input}
  A[-\mmaUnd{\mmaMu}]\,T[\mmaUnd{\mmaMu}]\,//\,GetArray
  First\,@\,GetArray[A[-\mmaUnd{\mmaMu}]]\,==\,%
\end{mmaCell}
\begin{mmaCell}{Output}
  \mmaPhi[t,x,y,z]
\end{mmaCell}
\begin{mmaCell}{Output}
  True
\end{mmaCell}

Using the partial derivative tensor operator $\partial_\mu$ we can calculate the electromagnetic tensor $F_{\mu\nu} = \partial_\mu A_\nu - \partial_\nu A_\mu$ and assign it to a tensor.
\begin{mmaCell}[moredefined={GetArray}]{Input}
  \mmaDef{\mmaEth}[-\mmaUnd{\mmaMu}][A[-\mmaUnd{\mmaNu}]] - \mmaDef{\mmaEth}[-\mmaUnd{\mmaNu}][A[-\mmaUnd{\mmaMu}]]\,//\,GetArray;
  F[-\mmaUnd{\mmaMu},-\mmaUnd{\mmaNu}]\,\mmaEquiv\,%
\end{mmaCell}
The antisymmetry of EM tensor can be shown either elementwise or by summing the arrays of normal and transposed arrays.
\begin{mmaCell}[addtoindex=1,moredefined={GetArray}]{Input}
  GetArray[F[-\mmaUnd{\mmaMu},-\mmaUnd{\mmaNu}]\,+\,F[-\mmaUnd{\mmaNu},-\mmaUnd{\mmaMu}]]
  GetArray[F[-\mmaUnd{\mmaMu},-\mmaUnd{\mmaNu}],\,\mmaCurly{-\mmaUnd{\mmaMu},-\mmaUnd{\mmaNu}}] + GetArray[F[-\mmaUnd{\mmaMu},-\mmaUnd{\mmaNu}],\,\mmaCurly{-\mmaUnd{\mmaNu},-\mmaUnd{\mmaMu}}]
\end{mmaCell}
\begin{mmaCell}{Output}
  \{\{0,0,0,0\},\{0,0,0,0\},\{0,0,0,0\},\{0,0,0,0\}\}
\end{mmaCell}
\begin{mmaCell}{Output}
  \{\{0,0,0,0\},\{0,0,0,0\},\{0,0,0,0\},\{0,0,0,0\}\}
\end{mmaCell}
From the electromagnetic tensor we can extract the vectors of the electric $E^\mu = F^{\mu\nu} T_\nu = (0, -\vec{\nabla}\phi - \partial_t\vec{A})$ and the magnetic fields $B^\mu = \frac{1}{2} \epsilon^{\mu\nu\lambda\rho} T_\nu F_{\lambda\rho} = (0, \vec{\nabla} \times \vec{A})$ (both purely spatial).
\begin{mmaCell}[moredefined={GetArray}]{Input}
  F[\mmaUnd{\mmaMu},\mmaUnd{\mmaNu}]\,T[-\mmaUnd{\mmaNu}]\,//\,GetArray\,//\,Column
  \mmaFrac{1}{2}\mmaUnd{\mmaEpsilon}[\mmaUnd{\mmaMu},\mmaUnd{\mmaNu},\mmaUnd{\mmaLambda},\mmaUnd{\mmaRho}]\,F[-\mmaUnd{\mmaLambda},-\mmaUnd{\mmaRho}]\,T[-\mmaUnd{\mmaNu}]\,//\,GetArray\,//\,Column
\end{mmaCell}
\begin{mmaCell}{Output}
  0
  -\mmaSup{\mmaPhi}{(0,1,0,0)}[t,x,y,z]-\mmaSup{A1}{(1,0,0,0)}[t,x,y,z]
  -\mmaSup{\mmaPhi}{(0,0,1,0)}[t,x,y,z]-\mmaSup{A2}{(1,0,0,0)}[t,x,y,z]
  -\mmaSup{\mmaPhi}{(0,0,0,1)}[t,x,y,z]-\mmaSup{A3}{(1,0,0,0)}[t,x,y,z]
\end{mmaCell}
\begin{mmaCell}{Output}
  0
  \mmaSup{A3}{(0,0,1,0)}[t,x,y,z]-\mmaSup{A2}{(0,0,0,1)}[t,x,y,z]
  \mmaSup{A1}{(0,0,0,1)}[t,x,y,z]-\mmaSup{A3}{(0,1,0,0)}[t,x,y,z]
  \mmaSup{A2}{(0,1,0,0)}[t,x,y,z]-\mmaSup{A1}{(0,0,1,0)}[t,x,y,z]
\end{mmaCell}

Let us now derive the field part of Maxwell equations.
First we define a general electromagnetic tensor using the electric and magnetic field components.
\begin{mmaCell}{Input}
  Ffield[-\mmaUnd{\mmaMu},-\mmaUnd{\mmaNu}]\,\mmaEquiv\,
  \mmaCurly{\mmaCurly{      0     ,\, Ex[t,x,y,z],\, Ey[t,x,y,z],\, Ez[t,x,y,z]},\\
   \mmaCurly{-Ex[t,x,y,z],\,      0     ,\,-Bz[t,x,y,z],\, By[t,x,y,z]},\\
   \mmaCurly{-Ey[t,x,y,z],\, Bz[t,x,y,z],\,      0     ,\,-Bx[t,x,y,z]},\\
   \mmaCurly{-Ez[t,x,y,z],\,-By[t,x,y,z],\, Bx[t,x,y,z],\,      0     }}
\end{mmaCell}
Next we calculate the field part of the homogeneous Maxwell equations $\partial_{\lambda}F_{\mu\nu}$ antisymmetrized by all indices.
For illustration we filter out trivial components.
This gives the well-known $\vec{\nabla}\times\vec{E} + \partial_t \vec{B} = 0$ and $\vec{\nabla}\cdot\vec{B} = 0$.
\begin{mmaCell}[moredefined={GetArray},morepattern={\#}]{Input}
  \mmaDef{\mmaEth}[-\mmaUnd{\mmaLambda}][Ffield[-\mmaUnd{\mmaMu},-\mmaUnd{\mmaNu}]] + \mmaDef{\mmaEth}[-\mmaUnd{\mmaMu}][Ffield[-\mmaUnd{\mmaNu},-\mmaUnd{\mmaLambda}]] + \mmaDef{\mmaEth}[-\mmaUnd{\mmaNu}][Ffield[-\mmaUnd{\mmaLambda},-\mmaUnd{\mmaMu}]]
  \hfill//\,GetArray\,//\,Flatten\,//\,DeleteCases[0]\,//\,DeleteDuplicatesBy[Sort[\mmaCurly{#,-#}]&]\,//\,Column
\end{mmaCell}
\begin{mmaCell}{Output}
   \mmaSup{Ex}{(0,0,1,0)}[t,x,y,z]-\mmaSup{Ey}{(0,1,0,0)}[t,x,y,z]-\mmaSup{Bz}{(1,0,0,0)}[t,x,y,z]
   \mmaSup{Ex}{(0,0,0,1)}[t,x,y,z]-\mmaSup{Ez}{(0,1,0,0)}[t,x,y,z]+\mmaSup{By}{(1,0,0,0)}[t,x,y,z]
   \mmaSup{Ey}{(0,0,0,1)}[t,x,y,z]-\mmaSup{Ez}{(0,0,1,0)}[t,x,y,z]-\mmaSup{Bx}{(1,0,0,0)}[t,x,y,z]
  -\mmaSup{Bz}{(0,0,0,1)}[t,x,y,z]-\mmaSup{By}{(0,0,1,0)}[t,x,y,z]-\mmaSup{Bx}{(0,1,0,0)}[t,x,y,z]
\end{mmaCell}
Finally, we calculate the field part of the inhomogeneous Maxwell equations $\partial_\mu F^{\mu\nu}$.
This gives the correct $\vec{\nabla}\cdot\vec{E}$ and $\vec{\nabla}\times\vec{B} - \partial_t \vec{E}$.
\begin{mmaCell}[moredefined={GetArray}]{Input}
  \mmaDef{\mmaEth}[-\mmaUnd{\mmaMu}][Ffield[\mmaUnd{\mmaMu},\mmaUnd{\mmaNu}]]\,//\,GetArray\,//\,Column
\end{mmaCell}
\begin{mmaCell}{Output}
   \mmaSup{Ez}{(0,0,0,1)}[t,x,y,z]+\mmaSup{Ey}{(0,0,1,0)}[t,x,y,z]+\mmaSup{Ex}{(0,1,0,0)}[t,x,y,z]
  -\mmaSup{By}{(0,0,0,1)}[t,x,y,z]+\mmaSup{Bz}{(0,0,1,0)}[t,x,y,z]-\mmaSup{Ex}{(1,0,0,0)}[t,x,y,z]
   \mmaSup{Bx}{(0,0,0,1)}[t,x,y,z]-\mmaSup{Bz}{(0,1,0,0)}[t,x,y,z]-\mmaSup{Ey}{(1,0,0,0)}[t,x,y,z]
  -\mmaSup{Bx}{(0,0,1,0)}[t,x,y,z]+\mmaSup{By}{(0,1,0,0)}[t,x,y,z]-\mmaSup{Ez}{(1,0,0,0)}[t,x,y,z]
\end{mmaCell}

The above calculation can also be done using the exterior calculus.
Let us obtain the electromagnetic potential 1-form $A$ by contracting the potential $A_\mu$ with the basis of 1-forms $dx^\mu$.
\begin{mmaCell}[moredefined={basis, GetArray}]{Input}
  Aform\;=\;A[-\mmaUnd{\mmaMu}]\,basis[\mmaUnd{\mmaMu}]\,//\,GetArray
\end{mmaCell}
\begin{mmaCell}{Output}
  dt\,\mmaPhi[t,x,y,z] - dx\,A1[t,x,y,z] - dy\,A2[t,x,y,z] - dz\,A3[t,x,y,z]
\end{mmaCell}
The electromagnetic field 2-form is the exterior derivative of the potential 1-form $F=dA$.
It is equal to the previously calculated electromagnetic tensor.
Note, that in our simple case we do not need to specify the space in which exterior derivative is taken.
\begin{mmaCell}[moredefined={Fform, ExteriorDerivative, Aform, basis, GetArray}]{Input}
  Fform\;=\;ExteriorDerivative[][Aform]\,//\,FullSimplify
  Fform\;==\;F[-\mmaUnd{\mmaMu},-\mmaUnd{\mmaNu}]\;*\;\mmaFrac{1}{2}\,basis[\mmaUnd{\mmaMu}]\mmaWedge\,\!basis[\mmaUnd{\mmaNu}]\,//\,GetArray\,//\,FullSimplify
\end{mmaCell}
\begin{mmaCell}{Output}
   dy\mmaWedge\,\!dz\;(\mmaSup{A2}{(0,0,0,1)}[t,x,y,z]\,-\,\mmaSup{A3}{(0,0,1,0)}[t,x,y,z]) 
  +dx\mmaWedge\,\!dy\;(\mmaSup{A1}{(0,0,1,0)}[t,x,y,z]\,-\,\mmaSup{A2}{(0,1,0,0)}[t,x,y,z])
  +dx\mmaWedge\,\!dz\;(\mmaSup{A1}{(0,0,0,1)}[t,x,y,z]\,-\,\mmaSup{A3}{(0,1,0,0)}[t,x,y,z])
  -dt\mmaWedge\,\!dx\;(\mmaSup{\mmaPhi}{(0,1,0,0)}[t,x,y,z]\,+\,\mmaSup{A1}{(1,0,0,0)}[t,x,y,z])
  -dt\mmaWedge\,\!dy\;(\mmaSup{\mmaPhi}{(0,0,1,0)}[t,x,y,z]\,+\,\mmaSup{A2}{(1,0,0,0)}[t,x,y,z])
  -dt\mmaWedge\,\!dz\;(\mmaSup{\mmaPhi}{(0,0,0,1)}[t,x,y,z]\,+\,\mmaSup{A3}{(1,0,0,0)}[t,x,y,z])
\end{mmaCell}
\begin{mmaCell}{Output}
  True
\end{mmaCell}
The homogeneous Maxwell equations are trivially satisfied $dF = ddA = 0$.
\begin{mmaCell}[moredefined={ExteriorDerivative}]{Input}
  ExteriorDerivative[][Fform]\,//\,FullSimplify
\end{mmaCell}
\begin{mmaCell}{Output}
  0
\end{mmaCell}
The field part of the inhomogeneous Maxwell equations $d(\star F)$ (for potentials) can be derived using an improvised Hodge star operator.
Although the result is cumbersome it is not hard to see that it reproduces the correct expression $\square A^\mu - \partial^\mu (\partial_\nu A^\nu)$.
\begin{mmaCell}[morepattern={f_, f, a_, b_, a, b},morelocal={tmp},moredefined={InteriorProduct, ExteriorDerivative}]{Input}
  HodgeStar[f_]\,\mmaEq Module[\mmaCurly{tmp},
  tmp\,=\,f\,/.\,\mmaCurly{a_\mmaWedge\,\!b_\,\mmaColonArrow a\mmaOtimes\,\!b\,-\,b\mmaOtimes\,\!a};
    tmp\,=\,tmp\,/.\,\mmaCurly{dt\,\mmaArrow\,\mmaUnd{\mmaEth\,\!t},\;dx\,\mmaArrow\,\mmaUnd{\mmaEth\,\!x},\;dy\,\mmaArrow\,\mmaUnd{\mmaEth\,\!y},\;dz\,\mmaArrow\,\mmaUnd{\mmaEth\,\!z}};
    \mmaFrac{1}{2}InteriorProduct[tmp,\;dt\mmaWedge\,\!dx\mmaWedge\,\!dy\mmaWedge\,\!dz]
  ];
  ExteriorDerivative[][HodgeStar[Fform]]\,//\,FullSimplify
\end{mmaCell}
\begin{mmaCell}[addtoindex=1]{Output}
  -dx\mmaWedge\,\!dy\mmaWedge\,\!dz\;(\mmaSup{\mmaPhi}{(0,2,0,0)}[t,x,y,z]\;+\;\mmaSup{\mmaPhi}{(0,0,2,0)}[t,x,y,z]\;+\;\mmaSup{\mmaPhi}{(0,0,0,2)}[t,x,y,z]
  \hfill+\;\mmaSup{A1}{(1,1,0,0)}[t,x,y,z]\;+\;\mmaSup{A2}{(1,0,1,0)}[t,x,y,z]\;+\;\mmaSup{A3}{(1,0,0,1)}[t,x,y,z])
  -dt\mmaWedge\,\!dy\mmaWedge\,\!dz\;(\mmaSup{A1}{(2,0,0,0)}[t,x,y,z]\;+\;\mmaSup{A1}{(0,0,2,0)}[t,x,y,z]\;+\;\mmaSup{A1}{(0,0,0,2)}[t,x,y,z]
  \hfill\;+\;\mmaSup{\mmaPhi}{(1,1,0,0)}[t,x,y,z]\;-\;\mmaSup{A2}{(0,1,1,0)}[t,x,y,z]\;-\;\mmaSup{A3}{(0,1,0,1)}[t,x,y,z])
  +dt\mmaWedge\,\!dx\mmaWedge\,\!dz\;(\mmaSup{A2}{(2,0,0,0)}[t,x,y,z]\;+\;\mmaSup{A2}{(0,2,0,0)}[t,x,y,z]\;+\;\mmaSup{A2}{(0,0,0,2)}[t,x,y,z]
  \hfill+\;\mmaSup{\mmaPhi}{(1,0,1,0)}[t,x,y,z]\;-\;\mmaSup{A1}{(0,1,1,0)}[t,x,y,z]\;-\;\mmaSup{A3}{(0,0,1,1)}[t,x,y,z])
  -dt\mmaWedge\,\!dx\mmaWedge\,\!dy\;(\mmaSup{A3}{(2,0,0,0)}[t,x,y,z]\;+\;\mmaSup{A3}{(0,2,0,0)}[t,x,y,z]\;+\;\mmaSup{A3}{(0,0,2,0)}[t,x,y,z]
  \hfill+\;\mmaSup{\mmaPhi}{(1,0,0,1)}[t,x,y,z]\;-\;\mmaSup{A1}{(0,1,0,1)}[t,x,y,z]\;-\;\mmaSup{A2}{(0,0,1,1)}[t,x,y,z])
\end{mmaCell}

Lastly, let us derive the field transformations under a boost.
First we define a general 4-dimensional space.
The indices of this space are \texttt{iMoving0}, \texttt{iMoving1}, etc..
\begin{mmaCell}[moredefined={DefineSpace}]{Input}
  DefineSpace["Moving",\;4]
\end{mmaCell}
Next we define the transformation matrix from the original space to the moving space with the help of a vielbein (tetrad) $e_\mu^i = \partial x'^i / \partial x^\mu$, where $x'^i = x'^i(x)$ is the coordinates of the moving frame.
Let the new system be moving in the z-direction with velocity $v$ in natural units.
\begin{mmaCell}[moredefined={metric}]{Input}
  metric[-\mmaUnd{\mmaMu},iMoving1]\;\mmaEquiv\;
  \mmaCurly{\mmaCurly{\mmaFrac{1}{\mmaSqrt{1-\mmaSup{v}{2}}},  0,  0,  \mmaFrac{-v}{\mmaSqrt{1-\mmaSup{v}{2}}}},\\
   \mmaCurly{   0,    1,  0,     0  },\\
   \mmaCurly{   0,    0,  1,     0  },\\
   \mmaCurly{\mmaFrac{-v}{\mmaSqrt{1-\mmaSup{v}{2}}},  0,  0,  \mmaFrac{1}{\mmaSqrt{1-\mmaSup{v}{2}}}}}
\end{mmaCell}
It is instructive to check that the moving space is also Minkowski.
Note, that this metric was not set manually, but calculated using the tetrad $g_{ij} = g_{\mu\nu} e^\mu_i e^\nu_j$.
\begin{mmaCell}[moredefined={metric, GetArray}]{Input}
  metric[-iMoving1,-iMoving2]\,//\,GetArray\,//\,Simplify\,//\,MatrixForm
\end{mmaCell}
\begin{mmaCell}[verbatimenv=]{Output}
  \hspace{1em}
   \(\begin{pmatrix}
     \texttt{1} & \texttt{0} & \texttt{0} & \texttt{0}\\
     \texttt{0} & \texttt{-1} & \texttt{0} & \texttt{0}\\
     \texttt{0} & \texttt{0} & \texttt{-1} & \texttt{0}\\
     \texttt{0} & \texttt{0} & \texttt{0} & \texttt{-1}
   \end{pmatrix}\)
\end{mmaCell}
The time direction vector in the moving space is:
\begin{mmaCell}{Input}
  Tmoving[iMoving0]\,\mmaEquiv\,\mmaCurly{1,0,0,0}
\end{mmaCell}
Now we can easily derive the well-known transformations for electric and magnetic fields in the moving frame using the same formulae as before.
\begin{mmaCell}[moredefined={GetArray}]{Input}
  Ffield[iMoving0,iMoving1]\,Tmoving[-iMoving1]\,//\,GetArray\,//\,FullSimplify
  \mmaFrac{1}{2}\mmaUnd{\mmaEpsilon}[iMoving0,iMoving1,iMoving2,iMoving3]\,Tmoving[-iMoving1]\,Ffield[-iMoving2,-iMoving3]
  \hfill//\,GetArray\,//\,FullSimplify
\end{mmaCell}
\begin{mmaCell}{Output}
  \{0, \mmaFrac{Ex[t,x,y,z]\;-\;v\,By[t,x,y,z]}{\mmaSqrt{1-\mmaSup{v}{2}}}, \mmaFrac{Ey[t,x,y,z]\;+\;v\,Bx[t,x,y,z]}{\mmaSqrt{1-\mmaSup{v}{2}}}, Ez[t,x,y,z]\}
\end{mmaCell}
\begin{mmaCell}{Output}
  \{0, \mmaFrac{Bx[t,x,y,z]\;+\;v\,Ey[t,x,y,z]}{\mmaSqrt{1-\mmaSup{v}{2}}}, \mmaFrac{By[t,x,y,z]\;-\;v\,Ex[t,x,y,z]}{\mmaSqrt{1-\mmaSup{v}{2}}}, Bz[t,x,y,z]\}
\end{mmaCell}
As a side note we mention, that \texttt{\mmaEth} is the partial derivative in the coordinates of its index, therefore the right-hand side of the first comparison below is in fact zero.
In expressions with basis changes, like $\partial_\mu F^{\mu\nu} = e_\mu^{\ a} \partial_a F^{\mu\nu} = \partial_a F^{a\mu}$, one must use \texttt{\mmaEth\textit{space}} to specify the coordinates explicitly, independent of the index.
\begin{mmaCell}[moredefined={GetArray, metric}]{Input}
  FullSimplify\,@\,GetArray[\mmaDef{\mmaEth}[-\mmaUnd{\mmaMu}][Ffield[\mmaUnd{\mmaMu},\mmaUnd{\mmaNu}]]]
  \hfill===\;FullSimplify\,@\,GetArray[metric[iMoving0,-\mmaUnd{\mmaMu}]\,\mmaDef{\mmaEth}[-iMoving0][Ffield[\mmaUnd{\mmaMu},\mmaUnd{\mmaNu}]]]
  FullSimplify\,@\,GetArray[\mmaDef{\mmaEth}[-\mmaUnd{\mmaMu}][Ffield[\mmaUnd{\mmaMu},\mmaUnd{\mmaNu}]]]
      ===\;FullSimplify\,@\,GetArray[metric[iMoving0,-\mmaUnd{\mmaMu}]\,\mmaUnd{\mmaEth\,\!Minkowski}[-iMoving0][Ffield[\mmaUnd{\mmaMu},\mmaUnd{\mmaNu}]]]
      ===\;FullSimplify\,@\,GetArray[\mmaUnd{\mmaEth\,\!Minkowski}[-iMoving0][Ffield[iMoving0,\mmaUnd{\mmaNu}]]]
\end{mmaCell}
\begin{mmaCell}{Output}
  False
\end{mmaCell}
\begin{mmaCell}{Output}
  True
\end{mmaCell}
Finally, when the space is no longer needed the following command effectively deletes it and all definitions connected to it.
\begin{mmaCell}[moredefined={DefineSpace}]{Input}
  DefineSpace["Moving",\;0]
  DefineSpace["Minkowski",\;0]
\end{mmaCell}

\subsection{Spinors}
In this part we will examine applications of SimpleTensor to Dirac and Weyl particles.
The following material can be found in the first few chapters of \cite{Peskin:1995ev}, although similar treatments are also given in many other quantum physics textbooks.
First, we will define the 4-vector of Dirac matrices and show how it can be manipulated into other matrices of the Clifford algebra using SimpleTensor.
We will solve the Dirac and Weyl equations, find the plane-wave solutions for both spinors and calculate the corresponding Berry phases \cite{Berry:1984jv}.
Lastly, we will check the transformations of Dirac spinors under a Lorentz boost.

We start again by defining a Minkowski space with mostly negative metric, the Levi-Civita tensor, and a time-direction 4-vector.
\begin{mmaCell}[moredefined={DefineSpace, metric}]{Input}
  DefineSpace["Minkowski",\;4,\;index\,\mmaArrow\,\mmaCurly{\mmaUnd{\mmaMu},\mmaUnd{\mmaNu},\mmaUnd{\mmaLambda},\mmaUnd{\mmaRho},\mmaUnd{\mmaAlpha},\mmaUnd{\mmaBeta},\mmaUnd{\mmaGamma},\mmaUnd{\mmaDelta}},\;coord\,\mmaArrow\,\mmaCurly{t,x,y,z}]
  metric[-\mmaUnd{\mmaMu},-\mmaUnd{\mmaNu}]\,\mmaEquiv\,DiagonalMatrix[\mmaCurly{1,-1,-1,-1}]
  \mmaUnd{\mmaEpsilon}[\mmaUnd{\mmaMu},\mmaUnd{\mmaNu},\mmaUnd{\mmaLambda},\mmaUnd{\mmaRho}]\,\mmaEquiv\,Normal\,@\,LeviCivitaTensor[4]
  T[\mmaUnd{\mmaMu}]\,\mmaEquiv\,\mmaCurly{1,0,0,0}
\end{mmaCell}
Next we define a 4-vector of gamma (Dirac) matrices $\gamma^\mu$ in chiral representation.
Note, that even though the list on the right-hand side has dimensions $\{4,4,4\}$ it is possible to assign it to a (matrix-valued) 4-vector $\gamma^\mu$ because the top-most dimension coincide with the Minkowski space dimension.
\begin{mmaCell}[verbatimenv=]{Input}
  \hspace{1em}\mmaUnd{\texttt{\mmaGamma}}[\mmaUnd{\texttt{\mmaMu}}]\,\mmaEquiv\,\big\{\(\begin{pmatrix}
     \texttt{0} & \texttt{0} & \texttt{1} & \texttt{0}\\
     \texttt{0} & \texttt{0} & \texttt{0} & \texttt{1}\\
     \texttt{1} & \texttt{0} & \texttt{0} & \texttt{0}\\
     \texttt{0} & \texttt{1} & \texttt{0} & \texttt{0}
   \end{pmatrix}\),
   \(\begin{pmatrix}
     \texttt{0} & \texttt{0} & \texttt{0} & \texttt{1}\\
     \texttt{0} & \texttt{0} & \texttt{1} & \texttt{0}\\
     \texttt{0} & \texttt{-1} & \texttt{0} & \texttt{0}\\
     \texttt{-1} & \texttt{0} & \texttt{0} & \texttt{0}
   \end{pmatrix}\),
   \(\begin{pmatrix}
     \texttt{0} & \texttt{0} & \texttt{0} & \texttt{-i}\\
     \texttt{0} & \texttt{0} & \texttt{i} & \texttt{0}\\
     \texttt{0} & \texttt{i} & \texttt{0} & \texttt{0}\\
     \texttt{-i} & \texttt{0} & \texttt{0} & \texttt{0}
   \end{pmatrix}\),
   \(\begin{pmatrix}
     \texttt{0} & \texttt{0} & \texttt{1} & \texttt{0}\\
     \texttt{0} & \texttt{0} & \texttt{0} & \texttt{-1}\\
     \texttt{-1} & \texttt{0} & \texttt{0} & \texttt{0}\\
     \texttt{0} & \texttt{1} & \texttt{0} & \texttt{0}
   \end{pmatrix}\)\big\}
\end{mmaCell}
It is easy to check, for example, that the trace of Gamma matrices product is proportional to the metric $\text{tr}(\gamma^\mu\gamma^\nu) = 4 g^{\mu\nu}$.
Note, that the \texttt{GetArray} command takes care of the dot product between the matrices as well as the trace operator.
\begin{mmaCell}[addtoindex=1,moredefined={metric, GetArray}]{Input}
  Tr[\mmaUnd{\mmaGamma}[\mmaUnd{\mmaMu}].\mmaUnd{\mmaGamma}[\mmaUnd{\mmaNu}]]\,==\,4\,metric[\mmaUnd{\mmaMu},\mmaUnd{\mmaNu}]\,//\,GetArray\,//\,FullSimplify
\end{mmaCell}
\begin{mmaCell}{Output}
  True
\end{mmaCell}
Using the 4-vector of gamma matrices it is possible to construct also the "fifth" gamma matrix $\gamma_5$ and a set of sigma matrices $\sigma^{\mu\nu}$.
Note, that in the chiral representation $\gamma_5$ is indeed diagonal and $P_{R/L}=(1\pm\gamma_5)/2$ are the projection operators onto the left- and right-handed components.
\begin{mmaCell}[moredefined={GetArray,i}]{Input}
  \mmaUnd{\mmaGamma\,\!5}\,=\,GetArray\(\bigg[\)\mmaFrac{-i}{24}\,\mmaUnd{\mmaEpsilon}[\mmaUnd{\mmaMu},\mmaUnd{\mmaNu},\mmaUnd{\mmaRho},\mmaUnd{\mmaLambda}]\,\mmaUnd{\mmaGamma}[-\mmaUnd{\mmaMu}].\mmaUnd{\mmaGamma}[-\mmaUnd{\mmaNu}].\mmaUnd{\mmaGamma}[-\mmaUnd{\mmaRho}].\mmaUnd{\mmaGamma}[-\mmaUnd{\mmaLambda}]\(\bigg]\); \mmaDef{\mmaGamma\,\!5}\,//\,MatrixForm
  \mmaUnd{\textit{\sigma}}[\mmaUnd{\mmaMu},\mmaUnd{\mmaNu}]\,\mmaEquiv\,GetArray\(\bigg[\)\mmaFrac{i}{2}\,(\mmaUnd{\mmaGamma}[\mmaUnd{\mmaMu}].\mmaUnd{\mmaGamma}[\mmaUnd{\mmaNu}]\;-\;\mmaUnd{\mmaGamma}[\mmaUnd{\mmaNu}].\mmaUnd{\mmaGamma}[\mmaUnd{\mmaMu}])\(\bigg]\);
\end{mmaCell}
\begin{mmaCell}[verbatimenv=]{Output}
  \hspace{1em}
   \(\begin{pmatrix}
     \texttt{-1} & \texttt{0} & \texttt{0} & \texttt{0}\\
     \texttt{0} & \texttt{-1} & \texttt{0} & \texttt{0}\\
     \texttt{0} & \texttt{0} & \texttt{1} & \texttt{0}\\
     \texttt{0} & \texttt{0} & \texttt{0} & \texttt{1}
   \end{pmatrix}\)
\end{mmaCell}
Let us also introduce for the following discussion a convenient conjugate-transpose (dagger) and an adjoint (bar) operations on spinors.
Note how we used $\gamma_0 = T_\mu\gamma^\mu$.
\begin{mmaCell}[addtoindex=3,moredefined={dagger, bar, GetArray},morepattern={a_, a, r_, i_, r, i},morelocal={res}]{Input}
  dagger[a_]\,\mmaEq Module[\mmaCurly{res},
    res\,=\,a\,/.\,Complex[r_,i_]\,\mmaColonArrow \,Complex[r,-i];
    If[MatchQ[Dimensions[res],\,\mmaCurly{x_,x_}],\;Transpose@res,\;res]
  ]
  bar[\mmaPat{\mmaPsi_}]\,:=\,dagger[\mmaPat{\mmaPsi}].GetArray[T[\mmaUnd{\mmaMu}]\,\mmaUnd{\mmaGamma}[-\mmaUnd{\mmaMu}]]
\end{mmaCell}

Next let us find a plane-wave solution $\psi$ to the Dirac equation for a particle with mass $m$ and 4-momentum $p^\mu$. 
First we define the 4-momentum and write the matrix part of the Dirac equation $(p_\mu \gamma^\mu - m)\psi = 0$.
Note that when adding a list-valued tensor to an explicit Mathematica list one should use the \texttt{TensorPlus} (\mmaOplus) operator.
  Otherwise Mathematica automatically threads the tensor inside the explicit list before any computation can take place yielding a list of lists in the result.
\begin{mmaCell}[moredefined={EqDirac, GetArray}]{Input}
  p[\mmaUnd{\mmaMu}]\,\mmaEquiv\,\mmaCurly{p0,p1,p2,p3}
  EqDirac\,=\,p[-\mmaUnd{\mmaMu}]\,\mmaUnd{\mmaGamma}[\mmaUnd{\mmaMu}] \mmaOplus -\,m\,IdentityMatrix[4]\,//\,GetArray;
  EqDirac\,//\,MatrixForm
\end{mmaCell}
\begin{mmaCell}[verbatimenv=]{Output}
  \hspace{1em}
   \(\begin{pmatrix}
     \texttt{-m} & \texttt{0} & \texttt{p0\;-\;p3} & \texttt{-p1\;+\;i\,p2}\\
     \texttt{0} & \texttt{-m} & \texttt{-p1\;-\;i\,p2} & \texttt{p0\;+\;p3}\\
     \texttt{p0\;+\;p3} & \texttt{p1\;-\;i\,p2} & \texttt{-m} & \texttt{0}\\
     \texttt{p1\;+\;i\,p2} & \texttt{p0\;-\;p3} & \texttt{0} & \texttt{-m}
   \end{pmatrix}\)
\end{mmaCell}
Let us now find the dispersion. 
A non-trivial solution exists only when the determinant is non-zero.
\begin{mmaCell}[moredefined={EqDirac},morefunctionlocal={p0}]{Input}
  Solve[Det[EqDirac]\,==\,0,\;p0]
\end{mmaCell}
\begin{mmaCell}{Output}
  \{\{p0\,\(\to\)\,-\,\mmaSqrt{\mmaSup{m}{2}+\mmaSup{p1}{2}+\mmaSup{p2}{2}+\mmaSup{p3}{2}}\}, \{p0\,\(\to\)\,-\,\mmaSqrt{\mmaSup{m}{2}+\mmaSup{p1}{2}+\mmaSup{p2}{2}+\mmaSup{p3}{2}}\},
       \{p0\,\(\to\)\,\mmaSqrt{\mmaSup{m}{2}+\mmaSup{p1}{2}+\mmaSup{p2}{2}+\mmaSup{p3}{2}}\}, \{p0\,\(\to\)\,\mmaSqrt{\mmaSup{m}{2}+\mmaSup{p1}{2}+\mmaSup{p2}{2}+\mmaSup{p3}{2}}\}\}
\end{mmaCell}
This is a doubly-degenerate dispersion for Dirac particles and anti-particles.
Let us choose the positive branch and find a conveniently normalized bi-spinor solution.
\begin{mmaCell}[moredefined={EqDirac}]{Input}
  EqDirac\,/.\,p0\,\mmaArrow\,\mmaSqrt{\mmaSup{m}{2}+\mmaSup{p1}{2}+\mmaSup{p2}{2}+\mmaSup{p3}{2}}\,//\,NullSpace
  \mmaUnd{\mmaPsi}\,=\,\mmaFrac{m}{\mmaSqrt{p3\,+\,\mmaSqrt{\mmaSup{m}{2}+\mmaSup{p1}{2}+\mmaSup{p2}{2}+\mmaSup{p3}{2}}}}\,First[
\end{mmaCell}
\begin{mmaCell}{Output}
  \{\{-\mmaFrac{p1\;-\;i\,p2}{m},\mmaFrac{p3\;+\;\mmaSqrt{\mmaSup{m}{2}+\mmaSup{p1}{2}+\mmaSup{p2}{2}+\mmaSup{p3}{2}}}{m},0,1\}, \{\mmaFrac{-p3\;+\;\mmaSqrt{\mmaSup{m}{2}+\mmaSup{p1}{2}+\mmaSup{p2}{2}+\mmaSup{p3}{2}}}{m},-\,\mmaFrac{p1\;+\;i\,p2}{m},1,0\}\}
\end{mmaCell}
We check that this spinor solution indeed has the current proportional to the momentum $j^\mu=\bar{\psi}\gamma^\mu\psi = 2p^\mu$ (factor 2 due to $\psi$ being a bi-spinor).
\begin{mmaCell}[addtoindex=1,moredefined={bar, GetArray}]{Input}
  bar[\mmaDef{\mmaPsi}].\mmaUnd{\mmaGamma}[\mmaUnd{\mmaMu}].\mmaDef{\mmaPsi}\,//\,GetArray\,//\,FullSimplify
\end{mmaCell}
\begin{mmaCell}{Output}
  \{2\,\mmaSqrt{\mmaSup{m}{2}+\mmaSup{p1}{2}+\mmaSup{p2}{2}+\mmaSup{p3}{2}},\;2\,p1,\;2\,p2,\;2\,p3\}
\end{mmaCell}
As a more advanced result, the Berry connection $\vec{a}_p = -i \bar{\psi}\vec{\nabla}_p\psi$ of the bi-spinor is zero because the Berry connections of comprising left and right spinors cancel each other $\vec{a}_p^R = -\vec{a}_p^L$.
\begin{mmaCell}[moredefined={bar},morepattern={\#}]{Input}
  -\mmaDef{i}\,bar[\mmaDef{\mmaPsi}].D[\mmaDef{\mmaPsi},\,#]\,&\,/@\,\mmaCurly{p1,p2,p3}\,//\,FullSimplify
  \mmaDef{\mmaPsi\,\!L}\,=\,\mmaFrac{1}{2}(IdentityMatrix[4]\,-\,\mmaDef{\mmaGamma\,\!5}).\mmaDef{\mmaPsi}; \mmaDef{\mmaPsi\,\!R}\,=\,\mmaFrac{1}{2}(IdentityMatrix[4]\,+\,\mmaDef{\mmaGamma\,\!5}).\mmaDef{\mmaPsi};
  bar[\mmaDef{\mmaPsi\,\!R}].D[\mmaDef{\mmaPsi\,\!L},#]\,==\,-\,bar[\mmaDef{\mmaPsi\,\!L}].D[\mmaDef{\mmaPsi\,\!R},\,#]\,&\,/@\,\mmaCurly{p1,p2,p3}\,//\,FullSimplify
\end{mmaCell}
\begin{mmaCell}{Output}
  \{0,0,0\}
\end{mmaCell}
\begin{mmaCell}[addtoindex=2]{Output}
  \{True,\,True,\,True\}
\end{mmaCell}

Let us now quickly calculate the Berry potential for a Weyl particle, as it will be needed in the next part.
The calculation is very similar to the one above.
First we define a 4-vector of Pauli matrices $\sigma^\mu$ and write the equation for a right-handed Weyl spinor $p_\mu\sigma^\mu\psi = 0$.
\begin{mmaCell}[morefunctionlocal={i},moredefined={EqWeyl, GetArray}]{Input}
  \mmaUnd{\mmaSigma}[\mmaUnd{\mmaMu}]\,\mmaEquiv\,Table[PauliMatrix[i],\,\mmaCurly{i,0,3}]
  EqWeyl\,=\,p[-\mmaUnd{\mmaMu}]\,\mmaUnd{\mmaSigma}[\mmaUnd{\mmaMu}]\,//\,GetArray;
  EqWeyl\,//\,MatrixForm
\end{mmaCell}
\begin{mmaCell}[verbatimenv=]{Output}
  \hspace{1em}
   \(\begin{pmatrix}
     \texttt{p0\;-\;p3} & \texttt{-p1\;+\;i\,p2} \\
     \texttt{-p1\;-\;i\,p2} & \texttt{p0\;+\;p3} 
   \end{pmatrix}\)
\end{mmaCell}
Next we find the dispersion.
\begin{mmaCell}[moredefined={EqWeyl},morefunctionlocal={p0}]{Input}
  Solve[Det[EqWeyl]\,==\,0,\;p0]
\end{mmaCell}
\begin{mmaCell}{Output}
  \{\{p0\,\(\to\)\,-\,\mmaSqrt{\mmaSup{p1}{2}+\mmaSup{p2}{2}+\mmaSup{p3}{2}}\},\;\{p0\,\(\to\)\,\mmaSqrt{\mmaSup{p1}{2}+\mmaSup{p2}{2}+\mmaSup{p3}{2}}\}\}
\end{mmaCell}
Now we find the right-handed Weyl spinor solution $u_R$ and normalize it to 1.
\begin{mmaCell}[moredefined={EqWeyl, uR}]{Input}
  EqWeyl\,/.\,p0\,\mmaArrow\,\mmaSqrt{\mmaSup{p1}{2}+\mmaSup{p2}{2}+\mmaSup{p3}{2}}\,//\,NullSpace
  uR\,=\,FullSimplify[Normalize\,@\,First[
\end{mmaCell}
\begin{mmaCell}{Output}
  \{\{\mmaFrac{p3\,+\,\mmaSqrt{\mmaSup{p1}{2}+\mmaSup{p2}{2}+\mmaSup{p3}{2}}}{p1\;+\;i\,p2},\;1\}\}
\end{mmaCell}
Finally we calculate the Berry potential $\vec{a}_p = -i \bar{u}_R\vec{\nabla}_p u_R$ and check that it indeed gives the monopole in momentum space at the coordinate origin $\vec{\nabla}\times\vec{a}_p = \frac{1}{2} \vec{p}/|p|^3$.
\begin{mmaCell}[addtoindex=1,moredefined={dagger, uR},morepattern={\#}]{Input}
  -\mmaDef{i}\,dagger[uR].D[uR,\,#]\,&\,/@\,\mmaCurly{p1,p2,p3}\,//\,FullSimplify
  Curl[
\end{mmaCell}
\begin{mmaCell}{Output}
  \{\mmaFrac{p2\,+\,\mmaFrac{p2\,p3}{\mmaSqrt{\mmaSup{p1}{2}+\mmaSup{p2}{2}+\mmaSup{p3}{2}}}}{2\,(\mmaSup{p1}{2}+\mmaSup{p2}{2})},\;-\,\mmaFrac{p1\,+\,\mmaFrac{p1\,p3}{\mmaSqrt{\mmaSup{p1}{2}+\mmaSup{p2}{2}+\mmaSup{p3}{2}}}}{2\,(\mmaSup{p1}{2}+\mmaSup{p2}{2})},\;0\}
\end{mmaCell}
\begin{mmaCell}{Output}
  \{\mmaFrac{p1}{2\,\mmaSup{(\mmaSup{p1}{2}+\mmaSup{p2}{2}+\mmaSup{p3}{2})}{3/2}},\;\mmaFrac{p2}{2\,\mmaSup{(\mmaSup{p1}{2}+\mmaSup{p2}{2}+\mmaSup{p3}{2})}{3/2}},\;\mmaFrac{p3}{2\,
\mmaSup{(\mmaSup{p1}{2}+\mmaSup{p2}{2}+\mmaSup{p3}{2})}{3/2}}\}
\end{mmaCell}

Let us now check that the Dirac spinor transforms correctly under Lorentz transformations.
For simplicity we consider a Lorentz boost in z-direction and parametrize it with an anti-symmetric tensor $w_{\mu\nu}$.
As we will check later $v$ is the speed of the moving frame.
\begin{mmaCell}[verbatimenv=]{Input}
  \hspace{1em}\mmaUnd{\texttt{w}}[-\mmaUnd{\texttt{\mmaMu}},-\mmaUnd{\texttt{\mmaNu}}]\,\mmaEquiv\,\(\begin{pmatrix}
     \texttt{0} & \texttt{0} & \texttt{0} & \texttt{Log}[\frac{1+\mmaUnd{\texttt{v}}}{\mmaSqrt{1-\mmaSup{\mmaUnd{v}}{2}}}]\\
     \texttt{0} & \texttt{0} & \texttt{0} & \texttt{0}\\
     \texttt{0} & \texttt{0} & \texttt{0} & \texttt{0}\\
     -\texttt{Log}[\frac{1+\mmaUnd{\texttt{v}}}{\mmaSqrt{1-\mmaSup{\mmaUnd{v}}{2}}}] & \texttt{0} & \texttt{0} & \texttt{0}
   \end{pmatrix}\)
\end{mmaCell}
Lorentz 4-vectors transform using the following representation $\Lambda_\text{vec}$ of the Lorentz group, where $J^{\mu\nu}_{\alpha\beta}$ is the transformation generator (see \cite{Peskin:1995ev} for details).
Note, that $p_\text{new}$ is indeed the momentum boosted with speed $v$ in z-direction.
\begin{mmaCell}[moredefined={GetArray, pnew}]{Input}
  J[\mmaUnd{\mmaMu},\mmaUnd{\mmaNu},-\mmaUnd{\mmaAlpha},-\mmaUnd{\mmaBeta}]\,\mmaEquiv\,Table[\mmaDef{i}\,(\mmaSub{\mmaDelta}{\mmaFnc{\mmaMu},\mmaFnc{\mmaAlpha}}\mmaSub{\mmaDelta}{\mmaFnc{\mmaNu},\mmaFnc{\mmaBeta}}\,-\,\mmaSub{\mmaDelta}{\mmaFnc{\mmaMu},\mmaFnc{\mmaBeta}}\mmaSub{\mmaDelta}{\mmaFnc{\mmaNu},\mmaFnc{\mmaAlpha}}),\,\mmaCurly{\mmaFnc{\mmaMu},0,3},\,\mmaCurly{\mmaFnc{\mmaNu},0,3},\,\mmaCurly{\mmaFnc{\mmaAlpha},0,3},\,\mmaCurly{\mmaFnc{\mmaBeta},0,3}]
  \mmaDef{{\Lambda}vec}\,=\,MatrixExp\,@\,GetArray[-\mmaFrac{\mmaDef{i}}{2}\,w[-\mmaUnd{\mmaMu},-\mmaUnd{\mmaNu}]\,J[\mmaUnd{\mmaMu},\mmaUnd{\mmaNu},\mmaUnd{\mmaAlpha},-\mmaUnd{\mmaBeta}]]\,//\,FullSimplify;
  pnew\,=\,\mmaDef{{\Lambda}vec}.GetArray[p[\mmaUnd{\mmaMu}]]\,//\,FullSimplify
\end{mmaCell}
\begin{mmaCell}[addtoindex=2]{Output}
  \{\mmaFrac{p0\;+\;p3\,v}{\mmaSqrt{1-\mmaSup{v}{2}}},\;p1,\;p2,\;\mmaFrac{p3\;+\;p0\,v}{\mmaSqrt{1-\mmaSup{v}{2}}}\}
\end{mmaCell}
Dirac spinors transform under $\Lambda_\text{sp}$ matrix representation with generators $\frac{1}{2}\sigma^{\mu\nu}$.
We check the result by confirming that the current of the transformed spinor is now proportional to the new momentum.
\begin{mmaCell}[moredefined={GetArray, bar, pnew}]{Input}
  \mmaDef{{\Lambda}sp}\,=\,MatrixExp[-\mmaFrac{\mmaDef{i}}{2}\,w[-\mmaUnd{\mmaMu},-\mmaUnd{\mmaNu}]\,*\,\mmaFrac{1}{2}\mmaUnd{\mmaSigma}[\mmaUnd{\mmaMu},\mmaUnd{\mmaNu}]\,//\,GetArray]\,//\,FullSimplify;
  \mmaDef{{\mmaPsi}new}\,=\,\mmaDef{{\Lambda}sp}.\mmaDef{{\mmaPsi}}\,//\,FullSimplify;
  bar[\mmaDef{{\mmaPsi}new}].\mmaUnd{\mmaGamma}[\mmaUnd{\mmaMu}].\mmaDef{{\mmaPsi}new}\,//\,GetArray\,//\,FullSimplify
\end{mmaCell}
\begin{mmaCell}[addtoindex=2]{Output}
  \{\mmaFrac{2\,(\mmaSqrt{\mmaSup{m}{2}+\mmaSup{p1}{2}+\mmaSup{p2}{2}+\mmaSup{p3}{2}}\;+\;v\,p3)}{\mmaSqrt{1-\mmaSup{v}{2}}},\;2\,p1,\;2\,p2,\;\mmaFrac{2\,(p3\;+\;v\mmaSqrt{\mmaSup{m}{2}+\mmaSup{p1}{2}+\mmaSup{p2}{2}+\mmaSup{p3}{2}})}{\mmaSqrt{1-\mmaSup{v}{2}}}\}
\end{mmaCell}
\begin{mmaCell}{Output}
  True
\end{mmaCell}
We can also check the transformation properties of the Dirac matrices directly $\Lambda_\text{sp}^{-1}\gamma^\mu\Lambda_\text{sp} = \Lambda^\mu_{\text{vec}\;\nu} \gamma^\nu$.
\begin{mmaCell}[moredefined={GetArray}]{Input}
  GetArray[Inverse[\mmaDef{{\Lambda}sp}].\mmaUnd{\mmaGamma}[\mmaUnd{\mmaMu}].\mmaDef{{\Lambda}sp}]\,==\,\mmaDef{{\Lambda}vec}.GetArray[\mmaUnd{\mmaGamma}[\mmaUnd{\mmaMu}]]\,//\,FullSimplify
\end{mmaCell}

\begin{mmaCell}{Output}
  True
\end{mmaCell}
Lastly we clear the Minkowski space to not interfere with the next part.
\begin{mmaCell}[moredefined={DefineSpace}]{Input}
  DefineSpace["Minkowski",\;0]
\end{mmaCell}

\subsection{Chiral Kinetic Theory}
In the last part we will examine applications of SimpleTensor to elementary differential geometric calculations on the example of chiral kinetic theory.
It is based mostly on the excellent papers by Stephanov, Yin, and others \cite{Stephanov:2012ki,Chen:2014cla}.
Note, that this material is more complicated and less widely known compared to previous sections.
Nevertheless, physics aside it illustrates how SimpleTensor allows calculations to closely follow the natural language of differential geometry.
First, we will write the Lagrangian 1-form for a chiral particle and check that it is Lorentz invariant.
Then we will construct the pre-symplectic form and find the evolution vector field for the chiral particle.
Finally, we will find the modification to the phase space volume and check the Liouville theorem for it.

In its simplest form chiral particles are living in the Euclidean phase space - a product of 3-dimensional spatial and momentum spaces, coupled to a 1-dimensional time line.
\begin{mmaCell}[moredefined={DefineSpace, metric}]{Input}
  DefineSpace["Spatial",\;3,\;index\,\mmaArrow\,\mmaCurly{i,j,k,l},\;coord\,\mmaArrow\,\mmaCurly{x,y,z}]
  DefineSpace["Momentum",\;3,\;index\,\mmaArrow\,\mmaCurly{a,b,c,d},\;coord\,\mmaArrow\,\mmaCurly{p1,p2,p3}]
  DefineSpace["Time",\;1,\;index\,\mmaArrow\,\mmaCurly{\mmaUnd{\mmaTau}},\;coord\,\mmaArrow\,\mmaCurly{t}]
  
  metric[-i,-j]\,\mmaEquiv\,IdentityMatrix[3]
  metric[-a,-b]\,\mmaEquiv\,IdentityMatrix[3]
\end{mmaCell}
Next we define in the spatial space a momentum vector $\vec{p}$ and in the momentum space a Berry potential vector $\vec{a}_p$ that was derived in the previous part.
\begin{mmaCell}[addtoindex=4]{Input}
  p[i]\,\mmaEquiv\,\mmaCurly{p1,p2,p3}
  ap[a]\,\mmaEquiv\,\(\Big\{\)\mmaFrac{p2\,+\,\mmaFrac{p2\,p3}{\mmaSqrt{\mmaSup{p1}{2}+\mmaSup{p2}{2}+\mmaSup{p3}{2}}}}{2\,(\mmaSup{p1}{2}+\mmaSup{p2}{2})},\;-\,\mmaFrac{p1\,+\,\mmaFrac{p1\,p3}{\mmaSqrt{\mmaSup{p1}{2}+\mmaSup{p2}{2}+\mmaSup{p3}{2}}}}{2\,(\mmaSup{p1}{2}+\mmaSup{p2}{2})},\;0\(\Big\}\)
\end{mmaCell}
Let us also introduce a uniform electric $\vec{E} = (0,0,E_z)$ and magnetic $\vec{B} = (0,0,B_z)$ fields both pointing in the z-direction.
We therefore define the following electromagnetic potential:
\begin{mmaCell}[addtoindex=1]{Input}
  \mmaUnd{\mmaVarphi}\,=\,Ez\,z;
  A[i]\,\mmaEquiv\,\mmaCurly{-\mmaFrac{1}{2}\,Bz\,y,\;\mmaFrac{1}{2}\,Bz\,x,\;0}
\end{mmaCell}
The dispersion $\epsilon = |\vec{p}| - \vec{B}\cdot\vec{s}$ of a relativistic chiral particle gets modified in the presence of the magnetic field due to its interaction with spin.
\begin{mmaCell}[addtoindex=1]{Input}
  \mmaUnd{\mmaEpsilon}\,=\,\mmaSqrt{\mmaSup{p1}{2}+\mmaSup{p2}{2}+\mmaSup{p3}{2}} - \mmaFrac{\mmaUnd{\mmaHbar}\,Bz\,p3}{2\,(\mmaSup{p1}{2}+\mmaSup{p2}{2}+\mmaSup{p3}{2})};
\end{mmaCell}
Now we have all ingredients to define the Lagrangian 1-form $L$ of a chiral particle.
Without the $\hbar$ correction this is exactly the usual classic-mechanical Lagrangian $L_\texttt{cl} = \epsilon - \varphi - (\vec{p} + \vec{A})\cdot\vec{v}$.
\begin{mmaCell}[moredefined={basis, GetArray}]{Input}
  L\,=\,(p[-i]\,+\,A[-i])\,basis[i] - (\mmaUnd{\mmaEpsilon}\,-\,\mmaUnd{\mmaVarphi})\,dt - \mmaUnd{\mmaHbar}\,ap[-a]\,basis[a]\,//\,GetArray;
\end{mmaCell}
Let us prove that this Lagrangian is indeed Lorentz invariant, at least up to the second order in $\hbar$.
For simplicity we consider a boost in the z-direction $\vec{\beta} = (0,0,\beta_3)$.
In this case the electric and magnetic fields do not change and therefore the Lagrangian is invariant with respect to the transformation if its Lie derivative in the direction of the vector field that generates this transformation is zero $\mathcal{L}_\delta L \approx 0$.
Under an infinitesimal Lorentz boost $\vec{\beta}$ the coordinates change by $\delta t = \vec{\beta}\cdot\vec{x}$, $\delta\vec{x} = t\vec{\beta}$ and $\delta p = \epsilon\vec{\beta}$.
This results in the following vector field for a regular particle $\delta = (\vec{\beta}\cdot\vec{x})\partial t + t(\vec{\beta}\cdot\partial\vec{x}) + \epsilon(\vec{\beta}\cdot\partial\vec{p})$.
In the case of the chiral particle this field gets an additional so-called side-jump term, i.e., position of a chiral particle depends on the Lorentz frame.
\begin{mmaCell}[morelocal={xvec, pvec, Bvec, jump}]{Input}
  With[\mmaCurly{\mmaLoc{\mmaBeta}\,=\,\mmaCurly{0,0,\mmaUnd{\mmaBeta3}},\;xvec\,=\,\mmaCurly{x,y,z},\;pvec\,=\,\mmaCurly{p1,p2,p3},\;Bvec\,=\,\mmaCurly{0,0,Bz}},
    With[\mmaCurly{jump\,=\,\mmaFrac{\mmaUnd{\mmaHbar}\,\mmaLoc{\mmaBeta}\mmaTimes\,\!pvec}{2\,(\mmaSup{p1}{2}+\mmaSup{p2}{2}+\mmaSup{p3}{2})}},

      \mmaUnd{\mmaDelta}\,=\,\mmaLoc{\mmaBeta}.xvec\;\mmaUnd{\mmaEth\,\!t} + (t\,\mmaLoc{\mmaBeta}\;+\;jump).\mmaCurly{\mmaUnd{\mmaEth\,\!x},\mmaUnd{\mmaEth\,\!y},\mmaUnd{\mmaEth\,\!z}} + (\mmaUnd{\mmaEpsilon}\,\mmaLoc{\mmaBeta}\;+\;jump\,\mmaTimes\,Bvec).\mmaCurly{\mmaUnd{\mmaEth\,\!p1},\mmaUnd{\mmaEth\,\!p2},\mmaUnd{\mmaEth\,\!p3}};
  ]]
\end{mmaCell}
Now we define the Lie derivative for differential forms using the Cartan formula $\mathcal{L}_X \omega = \iota_X d\omega + d(\iota_X \omega)$ and show that the Lagrangian is invariant (up to an irrelevant total derivative and up to $\hbar^2$) to the transformation generated by the boost $\delta$.
\begin{mmaCell}[addtoindex=-1,morepattern={v_, f_, v, f},moredefined={InteriorProduct, ExteriorDerivative}]{Input}
  LieDerivative[v_,\;f_]\,\mmaEq InteriorProduct[v,\;ExteriorDerivative[][f]]
  \hfill + ExteriorDerivative[][InteriorProduct[v,\;f]]

  \mmaDef{LieDerivative}[\mmaUnd{\mmaDelta},\;L] - ExteriorDerivative[]\(\Big[\)\mmaFrac{1}{2}\,Ez\,\mmaUnd{\mmaBeta3}\,(\mmaSup{t}{2}+\mmaSup{z}{2})\;+\;\mmaFrac{\mmaUnd{\mmaHbar}\,Bz\,\mmaUnd{\mmaBeta3\,(p1\,x\;+\;p2\,y)}}{4\,(\mmaSup{p1}{2}+\mmaSup{p2}{2}+\mmaSup{p3}{2})}\(\Big]\) + \mmaSup{O[\mmaUnd{\mmaHbar}]}{2}
\end{mmaCell}
\begin{mmaCell}[addtoindex=1]{Output}
  \mmaSup{O[\mmaHbar]}{2}
\end{mmaCell}

The Lagrangian 1-form (without the $dt$ term) works as a soldering form in our phase space and glues the particle velocity to its momentum.
The exterior derivative of a soldering form gives the symplectic form of the corresponding phase space.
However, our Lagrangian has an extra time coordinate and a term connected to it.
The exterior derivative $dL$ gives what is called a pre-symplectic form, i.e., an odd-dimensional 2-form whose null vector gives the system evolution and the quotient space with respect to this vector gives the phase-space.
The projection of $dL$ to this phase space is the symplectic form.
Let us therefore calculate the exterior derivative of $L$.
\begin{mmaCell}[moredefined={ExteriorDerivative}]{Input}
  dL\,=\,ExteriorDerivative[][L]\,//\,FullSimplify
\end{mmaCell}
\begin{mmaCell}{Output}
  dp1\mmaWedge\,\!dx + dp2\mmaWedge\,\!dy + dp3\mmaWedge\,\!dz - Ez\,dt\mmaWedge\,\!dz + Bz\,dx\mmaWedge\,\!dy - \mmaFrac{p1\,dp1\mmaWedge\,\!dt\;+\;p2\,dp2\mmaWedge\,\!dt\;+\;p3\,dp3\mmaWedge\,\!dt}{\mmaSqrt{\mmaSup{p1}{2}+\mmaSup{p2}{2}+\mmaSup{p3}{2}}}

  - \mmaHbar\(\Bigg[\)\mmaFrac{p1\,dp2\mmaWedge\,\!dp3\;-\;p2\,dp1\mmaWedge\,\!dp3\;+\;p3\,dp1\mmaWedge\,\!dp2}{2\,\mmaSup{(\mmaSup{p1}{2}+\mmaSup{p2}{2}+\mmaSup{p3}{2})}{3/2}} + \mmaFrac{Bz\,dp3\mmaWedge\,\!dt}{2\,(\mmaSup{p1}{2}+\mmaSup{p2}{2}+\mmaSup{p3}{2})}
  \hfill- \mmaFrac{Bz\,p3\,(p1\,dp1\mmaWedge\,\!dt\;+\;p2\,dp2\mmaWedge\,\!dt\;+\;p3\,dp3\mmaWedge\,\!dt)}{\mmaSup{(\mmaSup{p1}{2}+\mmaSup{p2}{2}+\mmaSup{p3}{2})}{2}}\(\Bigg]\)
\end{mmaCell}
\texttt{ExteriorDerivative} with empty first brackets is just a shortcut to take the derivative in the product space of all defined spaces.
\begin{mmaCell}[moredefined={ExteriorDerivative}]{Input}
  dL\,==\,ExteriorDerivative["Spatial",\;"Momentum",\;"Time"][L]
    \,==\,ExteriorDerivative["Spatial"][L] + ExteriorDerivative["Momentum"][L]
  \hfill + ExteriorDerivative["Time"][L]\,//\,FullSimplify
\end{mmaCell}
\begin{mmaCell}{Output}
  True
\end{mmaCell}
Let us now find the evolution equations.
We define a general vector field $v = \partial t + \vec{v}_x\cdot\partial\vec{x} + \vec{v}_p\cdot\partial\vec{p}$ and then demand it to be the null vector of $dL$, i.e., its contraction with must be zero $\iota_vdL = 0$.
The components of the form on the left-hand side are independent and therefore we get the system of evolution equations.
\begin{mmaCell}[moredefined={basis, GetArray, InteriorProduct}]{Input}
  vspatial[i]\,\mmaEquiv\,\mmaCurly{vx,vy,vz}
  vmomentum[a]\,\mmaEquiv\,\mmaCurly{vp1,vp2,vp3}
  v\,=\,\mmaUnd{\mmaEth\,\!t}\;+\;vspatial[i]\,basis[-i]\;+\;vmomentum[a]\,basis[-a]\,//\,GetArray
  EqEvolution\,=\,CoefficientArrays[InteriorProduct[v,\,dL],\;\mmaCurly{dt,dx,dy,dz,dp1,dp2,dp3}]
  \hfill\,//\,Last\,//\,Normal\,//\,FullSimplify
\end{mmaCell}
\begin{mmaCell}[addtoindex=2]{Output}
  \mmaEth\,\!t\;+\;\mmaEth\,\!p1\,vp1\;+\;\mmaEth\,\!p2\,vp2\;+\;\mmaEth\,\!p3\,vp3\;+\;\mmaEth\,\!x\,vx\;+\;\mmaEth\,\!y\,vy\;+\;\mmaEth\,\!z\,vz
\end{mmaCell}
\begin{mmaCell}{Output}
  -\,\mmaFrac{p1\,vp1\;+\;p2\,vp2\;+\;p3\,vp3}{\mmaSqrt{\mmaSup{p1}{2}+\mmaSup{p2}{2}+\mmaSup{p3}{2}}} + Ez\,vz + \mmaFrac{\mmaHbar\,Bz\,vp3}{2\,(\mmaSup{p1}{2}+\mmaSup{p2}{2}+\mmaSup{p3}{2})} - \mmaFrac{\mmaHbar\,Bz\,p3\,(p1\,vp1\;+\;p2\,vp2\;+\;p3\,vp3)}{\mmaSup{(\mmaSup{p1}{2}+\mmaSup{p2}{2}+\mmaSup{p3}{2})}{2}}

  vp1 - Bz\,vy

  vp2 + Bz\,vx

  vp3 - Ez

  -\,vx + \mmaFrac{p1}{\mmaSqrt{\mmaSup{p1}{2}+\mmaSup{p2}{2}+\mmaSup{p3}{2}}} + \mmaFrac{\mmaHbar\,Bz\,p1\,p3}{\mmaSup{(\mmaSup{p1}{2}+\mmaSup{p2}{2}+\mmaSup{p3}{2})}{2}} + \mmaFrac{\mmaHbar\,(p3\,vp2\;-\;p2\,vp3)}{2\,\mmaSup{(\mmaSup{p1}{2}+\mmaSup{p2}{2}+\mmaSup{p3}{2})}{3/2}}

  -\,vy + \mmaFrac{p2}{\mmaSqrt{\mmaSup{p1}{2}+\mmaSup{p2}{2}+\mmaSup{p3}{2}}} + \mmaFrac{\mmaHbar\,Bz\,p2\,p3}{\mmaSup{(\mmaSup{p1}{2}+\mmaSup{p2}{2}+\mmaSup{p3}{2})}{2}} + \mmaFrac{\mmaHbar\,(p1\,vp3\;-\;p3\,vp1)}{2\,\mmaSup{(\mmaSup{p1}{2}+\mmaSup{p2}{2}+\mmaSup{p3}{2})}{3/2}}

  -\,vz + \mmaFrac{p3}{\mmaSqrt{\mmaSup{p1}{2}+\mmaSup{p2}{2}+\mmaSup{p3}{2}}} - \mmaFrac{\mmaHbar\,Bz\,(\mmaSup{p1}{2}+\mmaSup{p2}{2}-\mmaSup{p3}{2})}{2\,\mmaSup{(\mmaSup{p1}{2}+\mmaSup{p2}{2}+\mmaSup{p3}{2})}{2}} - \mmaFrac{\mmaHbar\,(p1\,vp2\;-\;p2\,vp1)}{2\,\mmaSup{(\mmaSup{p1}{2}+\mmaSup{p2}{2}+\mmaSup{p3}{2})}{3/2}}
\end{mmaCell}
Note, that disregarding $\hbar$ corrections the above equations amount to $\dot{x} = \vec{p}/|\vec{p}|$, i.e., a particle moving with a speed of light in the direction of its momentum, and $\dot{p} = \vec{E} + \vec{v}\times\vec{B}$, i.e., the Lorentz force.
This system of equations indeed has the unique solution for the system evolution.
The resulting vector $v_\texttt{evol}$, however, is quite cumbersome, its value can be found in the SimpleTensor example notebook. 
\begin{mmaCell}[morefunctionlocal={vx, vy, vz, vp1, vp2, vp3}]{Input}
  vEvol\,=\,v\,/.\,First\,@\,Solve[EqEvolution\,==\,0,\;\mmaCurly{vx,vy,vz,vp1,vp2,vp3}]\,//\,Normal\,//\,FullSimplify;
\end{mmaCell}
Due to complexity of the evolution it is hard to actually find the canonical coordinates of the phase space and project $dL$ onto them.
It is still easy, however, to find the phase space volume, as it can be found from the highest non-zero power of the symplectic form.
Note the $\hbar$ correction in the brackets, its origin is intimately connected to the Berry potential $\vec{a}_p$ and the chiral anomaly.
\begin{mmaCell}{Input}
  vol\,=\,\mmaFrac{1}{3!}\,dL\mmaWedge\,\!dL\mmaWedge\,\!dL\mmaWedge\,\!dt\,//\,FullSimplify
\end{mmaCell}
\begin{mmaCell}{Output}
  \(\bigg(\)1\;+\;\mmaFrac{\mmaHbar\,Bz\,p3}{2\,\mmaSup{(\mmaSup{p1}{2}+\mmaSup{p2}{2}+\mmaSup{p3}{2})}{3/2}}\(\bigg)\)\,dp1\mmaWedge\,\!dp2\mmaWedge\,\!dp3\mmaWedge\,\!dt\mmaWedge\,\!dx\mmaWedge\,\!dy\mmaWedge\,\!dz
\end{mmaCell}
Let us finally check that the Liouville theorem in this theory still holds and the volume form stays invariant during the evolution.
In other words, we show that its Lie derivative in the direction of evolution is zero.
\begin{mmaCell}{Input}
  LieDerivative[vEvol,\;vol]\,//\,FullSimplify
\end{mmaCell}
\begin{mmaCell}{Output}
  0
\end{mmaCell}
Lastly, after all the calculations we tidy up after ourselves.
\begin{mmaCell}[moredefined={DefineSpace}]{Input}
  DefineSpace["Spatial",\;0]
  DefineSpace["Momentum",\;0]
  DefineSpace["Time",\;0]
\end{mmaCell}

\section{Summary}
SimpleTensor is the new Mathematica package for componentwise tensor and differential-geometric calculations.
Among its advantages (or at least distinctions) over alternatives are the ease of use, conciseness, flexibility, and hackability.
It provides the most straightforward route to the result with minimal boilerplate code and is designed to be easily accessible.
It contains only a handful of functions on top of which a user can build up depending on her or his needs, as we have done in the examples with \texttt{LieDerivative} or \texttt{HodgeStar}.
Moreover, the codebase itself is short, documented, and supplemented by tests, which enables correction to the package itself, if need arises.
Note, that due to deliberate lack of optimisation in favor of code clarity SimpleTensor is slower compared to the alternatives and is not suited for heavy calculations.
SimpleTensor is also not specialized, providing only essential functionality, and so it lacks many frequently used functions for general relativity or quantum field theory.
Despite these limitations, as illustrated in the examples, it can be used quite easily in settings ranging from an undergraduate level electrodynamics to relativistic quantum mechanics to even actively researched chiral kinetic theory.

\end{document}